%% file: manuscript.tex
\documentclass[11pt]{article}
\usepackage[letterpaper,margin=1in]{geometry}
\usepackage{amsmath,amssymb,amsthm,graphicx,booktabs,natbib}
\usepackage{longtable,array,caption}
\usepackage[section]{placeins}
\usepackage[colorlinks=true,allcolors=black]{hyperref}
\newtheorem{proposition}{Proposition}

\newcommand{\R}{\mathbb R}
\newcommand{\bbeta}{\beta}
\newcommand{\soft}{\mathcal S}
\newcommand{\supp}{\operatorname{supp}}
\newcommand{\sgn}{\operatorname{sign}}
\newcommand{\Lloc}{\mathcal L}
\title{\bfseries Local Epochs, Averaging, and Variable Selection\\in Federated Lasso}
\ifdefined\anonymous
\author{}
\else
\author{Keivan Bolouri\\[2pt]
\small Department of Statistics, Donald Bren School of Information and Computer Sciences\\
\small University of California, Irvine\\
\small Corresponding author: \texttt{kbolouri@uci.edu}}
\fi
\date{}
\begin{document}
\maketitle
\input{sections/abstract}
\input{sections/introduction}
\input{sections/setup}
\input{sections/theory}
\input{sections/methods}
\input{sections/design}
\input{sections/results}
\input{sections/discussion}
\section*{Data and code availability}
The code and simulation files used for this paper are available in the project
repository, including the files needed to reproduce the tables and figures:
\url{https://github.com/KeivanBolouri/federated-lasso}.
\section*{Supplementary material}
The supplement provides the proof of the correlated example, the full simulation
grid, paired comparisons, threshold sensitivity, resource accounting, and context
from the earlier experiment with site-specific penalties.
\section*{Funding}
This research received no dedicated funding.
\section*{Disclosure statement}
The author reports no conflict of interest.
\bibliography{refs}
\end{document}

%% file: sections/abstract.tex
\begin{abstract}
\noindent\textit{\{Theoretical analysis and Monte Carlo simulation separate
local-epoch effects from averaging and tuning effects in federated Lasso.\}}

How much local work should precede averaging when fitting a sparse regression?
An orthogonal calculation shows that extra epochs can have no effect while
averaging still enlarges the selected set. A correlated two-site construction
gives an exact, nonmonotone limiting objective gap and its minimizing epoch
count. We then compare coordinate-descent averaging, two thresholding
modifications, and adapted FedDualAvg across twelve scenarios and 600
replicates. Methods share a penalty, independent validation samples, selection
rules, and resource limits. FedDualAvg generally achieves smaller objective gaps but
does not always recover variables better. Thresholding gains depend strongly
on selection rules. Epoch effects vary with correlation, signal strength,
site allocation, and the outcome measured. These results distinguish faster
iteration from better optimization, prediction, and variable selection.
\end{abstract}
\noindent\textbf{Keywords:} federated learning; Lasso; coordinate descent;
local epochs; variable selection; dual averaging.

%% file: sections/introduction.tex
\section{Introduction}
\label{sec:intro}

Federated learning allows several sites to fit a model while keeping their
individual records at each site. A server sends a model to the sites,
each site updates it using its own data, and the server combines the
updates \citep{mcmahan2017,kairouz2021}. One practical choice is how much
work a site should do before sending an update. More local work can reduce
the number of communication rounds. It can also move the local models
farther apart when sites have different data
\citep{stich2019,li2020convergence,karimireddy2020}.

This choice has an additional consequence for the Lasso, which estimates
a regression model while setting some coefficients to zero
\citep{tibshirani1996}. The nonzero coefficients identify the selected
variables. Coordinate descent fits the Lasso by updating one coefficient
at a time \citep{friedman2010}. A simple federated implementation lets each
site perform several complete sweeps through the coefficients and then
averages the resulting vectors. We call each sweep a local epoch and
write $E$ for the number of epochs between averages.

Two separate questions arise. First, how does $E$ affect the model reached
by this procedure? Second, does averaging preserve the variable selection
achieved by the local Lasso fits? The questions are related, but they are
not interchangeable. A variable selected at just one site can remain
nonzero after averaging, although its averaged coefficient is small.
Increasing the number of local epochs does not directly remove that
coefficient. With correlated predictors, changing $E$ can also change the
local update itself and therefore change the model produced by repeated
averaging.

The difficulty of averaging sparse coefficient vectors is established in
the federated composite-optimization literature. \citet{yuan2021fco}
analyzed this problem and introduced federated dual averaging
(FedDualAvg), which combines dual information before applying the
regularization step. They also evaluated sparse recovery.
\citet{bao2022} developed further dual-averaging methods with optimization
and statistical-recovery guarantees. These papers provide the starting
point for our study. Local fixed-point methods and analyses of local-update bias also provide
relevant precedents \citep{malinovsky2020,wang2024on}. Our exact examples
illustrate these issues for cyclic coordinate descent; they do not claim
a new general theory of local updates.

Our aim is to explain the role of local epochs in a particular cyclic
coordinate-descent procedure and to evaluate possible changes to that
procedure against an established alternative. We consider ordinary
coefficient averaging, soft-thresholding once after fitting (ST),
soft-thresholding after every average (P), and an adapted FedDualAvg
implementation. ST and P are simple modifications whose value must be
measured; neither is assumed to solve the pooled Lasso problem.

The article makes three specific contributions. First, it separates the
effect of local epochs from the effect of averaging. An exact orthogonal
calculation shows a setting in which additional epochs have no effect,
although averaging can still enlarge the selected set. A correlated
two-site example then shows that the limiting objective error can vary
nonmonotonically with $E$. Together, these calculations explain why
performing more local optimization need not improve the common
objective.

Second, we compare all four procedures using a common Lasso penalty,
the same generated datasets, the same validation rule, and common
resource limits. We examine an upload budget and a joint budget for
uploads and leading arithmetic work. Upload totals include every fitted path and the scalar validation
exchanges used for tuning. The arithmetic limit concerns leading training work;
setup and validation times are reported separately. This comparison addresses a
practical question: after paying for selection of its tuning parameters,
does a thresholding modification offer a useful trade-off relative to
ordinary averaging and FedDualAvg?

Third, targeted simulation scenarios vary predictor correlation, site
count, signal strength, predictor dimension, and site-size imbalance to
examine when the epoch choice matters. We measure objective error,
prediction error, support recovery,
communication, and local work separately. These outcomes answer different
questions. A model can be closer to the pooled Lasso solution without
identifying more true variables; conversely, additional thresholding can
improve variable selection while worsening prediction. Reporting the
outcomes together makes these trade-offs visible.

Related statistical work studies communication-efficient sparse
regression and distributed inference
\citep{lee2017,battey2018,jordan2019,wang2017}. Consensus methods offer
another way to solve a shared penalized objective \citep{boyd2011}, while
proximal methods handle nonsmooth regularization more broadly
\citep{parikh2014,tran2021}. Our focus is narrower: understanding the
local-epoch choice for coordinate-descent averaging and assessing its
thresholding modifications under explicit tuning and cost conventions.
An earlier experiment with separately selected site penalties is retained
as supplementary context. The main comparison uses a common penalty so
that differences between procedures have a clearer interpretation.

%% file: sections/setup.tex
\section{The common objective and local updates}
\label{sec:setup}

\subsection{What the sites are trying to fit}

There are $K$ sites and $p$ predictors. Site $j$ has a predictor matrix
$X_j\in\R^{m_j\times p}$ and response vector $y_j\in\R^{m_j}$.
The total training sample size is $m=\sum_{j=1}^K m_j$. We give each site
weight $w_j=m_j/m$, so each training observation has the same weight in
the combined loss. Predictor columns have positive squared norm at every
site. All procedures in the main comparison use the same penalty
$\lambda>0$, held fixed during fitting.

The local Lasso objective is
\begin{equation}
\Lloc_j(\bbeta)
 =\frac{1}{2m_j}\|y_j-X_j\bbeta\|_2^2
  +\lambda\|\bbeta\|_1,
\qquad \bbeta\in\R^p.
\label{eq:local}
\end{equation}
The first term measures the local fit; the second penalizes coefficient
size and can set coefficients to zero. The sample-size normalization
places each local loss on the same scale. Our common objective is
\begin{equation}
F(\bbeta)=\sum_{j=1}^K w_j\Lloc_j(\bbeta)
 =\frac{1}{2m}\|y-X\bbeta\|_2^2+\lambda\|\bbeta\|_1,
\label{eq:global}
\end{equation}
where $X$ and $y$ denote the vertically concatenated data. The last equality
follows from $w_j/m_j=1/m$ and $\sum_jw_j=1$. Concatenation defines the
reference objective; the federated procedure does not require records to
be transferred to the server.

Write $\hat\bbeta_F$ for a minimizer of $F$. The objective gap
$F(\bbeta)-F(\hat\bbeta_F)$ measures how far a fitted model is from the
minimum objective value. It does not measure whether the selected
variables are correct. We write $\supp(\bbeta)=\{k:\bbeta_k\ne0\}$ for
the mathematical support. Numerical support assessments use the explicit
activity threshold stated with the experiments.

\subsection{One local epoch and one communication round}

Coordinate descent updates one coefficient while holding the others
fixed. Define soft-thresholding by
\[
\soft(v,a)_k=\sgn(v_k)\max\{|v_k|-a,0\},\qquad a\ge0.
\]
It shrinks each coefficient toward zero and removes those whose magnitude
does not exceed $a$ \citep{parikh2014}. At site $j$, let $x_{jk}$ be
predictor column $k$ and let
$r=y_j-X_j\bbeta+x_{jk}\bbeta_k$ be the residual with the current
contribution of that column removed. The exact local update is
\begin{equation}
\bbeta_k\leftarrow
\frac{\soft(x_{jk}^{\top}r,m_j\lambda)}{\|x_{jk}\|_2^2}.
\label{eq:cd}
\end{equation}
The threshold contains $m_j$ because the loss in~\eqref{eq:local}
is divided by $m_j$. One epoch applies this update to coordinates
$1,\ldots,p$ in that fixed order. Let $T_j^E$ denote $E$ such epochs.

The ordinary averaging procedure starts at $\bbeta_0=0$. In round $t$,
the server sends $\bbeta_t$ to every site. Each site starts from this
same vector, performs $E$ local epochs, and returns its updated
coefficients. The server then computes
\begin{equation}
\bbeta_{t+1}=\sum_{j=1}^K w_jT_j^E(\bbeta_t).
\label{eq:average}
\end{equation}
The experiments specify the available number of rounds and any stopping
rule. Reaching a small change between successive averages does not
establish that $F$ has been minimized.

A single path of $R$ rounds uploads $8RKp$ bytes when coefficients are
sent as dense double-precision vectors. It uses $RE$ coordinate-descent
epochs at each site. Additional fitted paths and validation exchanges
used for tuning contribute to the reported total cost. These upload
counts exclude downloads and network overhead. Local work is reported
separately because reducing communication rounds can require more
computation at the sites.

%% file: sections/theory.tex
\section{What local epochs can and cannot change}
\label{sec:theory}

\subsection{An orthogonal design separates epochs from sparsity}

Suppose $X_j^\top X_j=m_jI_p$ at every site, and write
$z_j=X_j^\top y_j/m_j$. This assumption requires $p\leq\min_jm_j$.
Each coordinate update then ignores the other coefficients. One complete
sweep gives the local solution, regardless of the starting point:
\[
 T_j^E(b)=\soft(z_j,\lambda)\quad\text{for every }E\geq1.
\]
Consequently, all epoch counts give the same server estimate
$\bar b=\sum_jw_j\soft(z_j,\lambda)$ after the first round.
The pooled Lasso solution is
$\hat\bbeta_F=\soft(\sum_jw_jz_j,\lambda)$.
Thresholding before averaging and thresholding after averaging need not give
the same answer.

For fixed penalties and weights, the support of $\bar b$ is the union of
the local supports, except when nonzero contributions cancel exactly.
To see the qualification, fix the active coordinates and their signs at
each site. Each averaged coefficient is then an affine function of the
local scores. If any site selects that coefficient, its score has nonzero
weight, so exact cancellation lies on a proper hyperplane. The exceptional
set has Lebesgue measure zero. Also, a coefficient absent from every local
fit has $|z_{jk}|\leq\lambda$ at every site, which implies
$|\sum_jw_jz_{jk}|\leq\lambda$. Thus the pooled support is contained in
the union of the local supports.

This calculation identifies an averaging effect that extra epochs cannot
repair. It does not say that all averages are dense. Nor does it apply
directly to high-dimensional local designs. The next example addresses
correlation explicitly.

\subsection{With correlation, faster convergence per round can lead to a worse limit}

Consider two equally weighted sites and two predictors. Their normalized
Gram matrices are
\[
 H_1=I_2,\qquad H_2=\begin{pmatrix}1&r\\r&1\end{pmatrix},\qquad 0<r<1.
\]
Choose $a,d,\lambda>0$, let $\theta_1=(a,a)^\top$ and
$\theta_2=(a,a+d)^\top$, and set the score vectors to
$z_j=H_j\theta_j+\lambda(1,1)^\top$.
These are valid least-squares problems with positive-definite Gram
matrices. The two local Lasso solutions are $\theta_1$ and $\theta_2$.
All sweeps update coordinate 1 before coordinate 2.

\begin{proposition}[The limit depends on the epoch count]
\label{prop:correlated}
Starting from zero, the averaging iteration converges to
\begin{equation}
 b_E=\begin{pmatrix}
 a+\dfrac{r^{2E-1}d}{2(2-r^{2E})}\\[5pt]
 a+\dfrac{(1-r^{2E})d}{2-r^{2E}}
 \end{pmatrix}.
 \label{eq:limit-main}
\end{equation}
The second-coordinate error contracts by $r^{2E}/2$ in each round.
The pooled solution is
$b_\star=(a+rd/(4-r^2),\ a+(2-r^2)d/(4-r^2))^\top$.
Writing $t_E=r^{2E}/(2-r^{2E})$, the exact objective gap is
\begin{equation}
 F(b_E)-F(b_\star)
 =\frac{d^2r^4}{32(4-r^2)}
  +\frac{d^2}{8r^2}\left(t_E-\frac{r^2}{2}\right)^2.
 \label{eq:gap-main}
\end{equation}
All local updated coefficients and all noninitial server coefficients
are positive.
\end{proposition}

The proof is in Supplement S1. Its main step is simple: if the incoming
second coefficient at site 2 is $u\in[0,a+d]$, one sweep changes it to
$(1-r^2)(a+d)+r^2u$. This interval is preserved by further sweeps and
server averaging. Repeating the update $E$ times gives
\eqref{eq:limit-main}; the positive coefficients make the objective-gap
calculation quadratic.

Equation~\eqref{eq:gap-main} has two implications. First, its positive
constant term means that no choice of $E$ removes the averaging error in
this example. Second, its variable term can decrease and then increase
as $E$ grows. Extending the formula to real epoch counts gives the unique
minimum at
\begin{equation}
 E_\star(r)=1+\frac{\log(1+r^2/2)}{-\log(r^2)}.
 \label{eq:best-main}
\end{equation}
This value increases with $r$. The best integer choice is found by
comparing the two adjacent integers in \eqref{eq:gap-main}.

\begin{figure}[htbp]
\centering
\includegraphics[width=\textwidth]{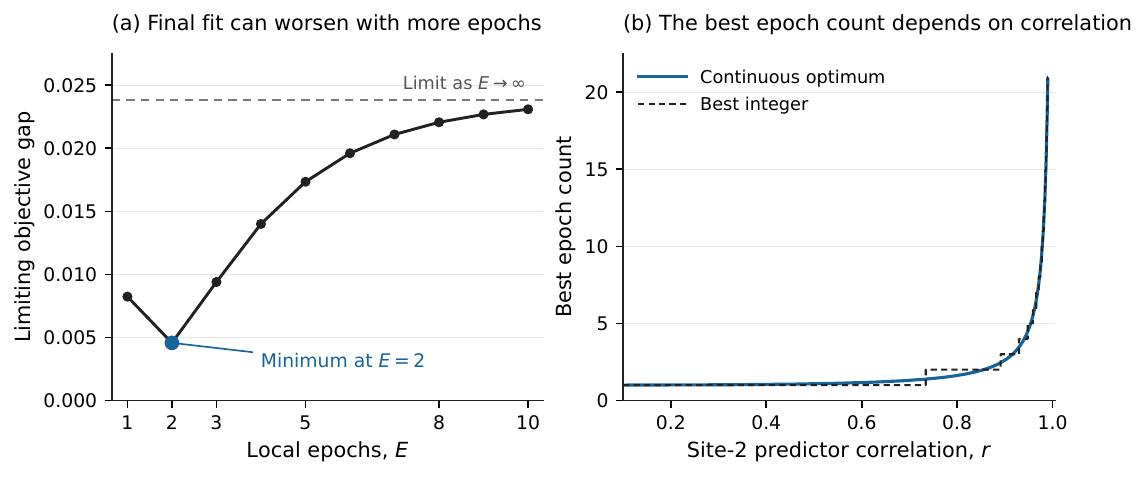}
\caption{Exact calculations for the correlated two-site construction.
The left panel shows the limiting objective gap at $r=0.8$, $a=2$, and
$d=1$. The right panel shows how its minimizing epoch count changes with
$r$. These are deterministic formulas, not simulation averages.}
\label{fig:theory}
\end{figure}

For example, at $r=0.8$, $a=2$, and $d=1$, the objective gaps for
$E=1,2,3$ are $0.008239$, $0.004571$, and $0.009398$.
Two epochs give the smallest of these gaps. Meanwhile, the round-by-round
contraction factors improve from $0.3200$ to $0.2048$ to $0.1311$.
Thus faster convergence of the averaging iteration does not imply a
better limiting fit. Figure~\ref{fig:theory} displays the calculation.

The example isolates optimization error while the selected variables
remain unchanged. Its optimal epoch count concerns a limiting objective
and the stated coordinate order. It is not a rule for choosing $E$ in
other datasets or under a finite budget. We next evaluate finite-budget
fits and support recovery directly.

%% file: sections/methods.tex
\section{Four procedures under the same comparison rules}
\label{sec:methods}

All four procedures start from zero and use the common penalty in
\eqref{eq:global}. They receive the same training and validation data in each
replicate. We keep the short names CD, ST, P, and DA throughout the results.

\paragraph{CD: average local coordinate-descent fits.}
This is the update in \eqref{eq:average}. The server returns its last available
iterate. Its candidate set contains one model for a specified epoch count and
budget.

\paragraph{ST: threshold the final CD average.}
ST first runs CD. It then considers five versions
$\soft(\bbeta_R,\tau)$, with
$\tau/\lambda\in\{0,0.03,0.1,0.3,1\}$.
The threshold is selected using validation data. All five candidates share the
same training path, so ST adds validation and thresholding work but does not need
five training runs. The zero threshold includes the unmodified CD estimate.

\paragraph{P: threshold after every average.}
P replaces the server update with
\[
 \bbeta_{t+1}=\soft\!\left(\sum_j w_jT_j^E(\bbeta_t),\tau\right).
\]
It uses the same five thresholds as ST. Each threshold changes later local
updates, so each candidate requires its own complete path. We count all five
paths. This procedure is a repeated sparsification of CD averaging; it is not
assumed to minimize $F$.

\paragraph{DA: adapted federated dual averaging.}
DA implements the full-participation, full-gradient specialization of
FedDualAvg in \citet{yuan2021fco}, with sample-size weights, squared Euclidean
geometry, and server learning rate one. At zero-based round $t$ and local step
$k$, site $j$ forms
\[
 b_{j,t,k}=\soft\{z_{j,t,k},\eta(tE+k)\lambda\},\qquad
 z_{j,t,k+1}=z_{j,t,k}-\eta\nabla f_j(b_{j,t,k}),
\]
where $f_j(b)=\|y_j-X_jb\|_2^2/(2m_j)$. The server averages the final dual
vectors and applies threshold $\eta(t+1)E\lambda$ to obtain the next model.
Each round starts with $z_{j,t,0}=z_t$, and the server sets
$z_{t+1}=\sum_jw_jz_{j,t,E}$, starting from $z_0=0$.
The five candidate client rates are
$\eta\in\{0.03,0.1,0.3,0.6,1\}/L_{\max}$, where
$L_{\max}=\max_j\lambda_{\max}(X_j^\top X_j/m_j)$.
The largest local eigenvalue is computed at each site and communicated as a
scalar. Every rate defines a separate path and all five paths are charged.
We return the final server model for the selected rate. Each gradient is
evaluated directly from the local design matrix.

\subsection{The same rule chooses each method's candidate}

The primary rule selects the candidate with the smallest pooled validation
mean squared error. A secondary rule selects the candidate with the fewest
active coefficients among those within one estimated standard error of the
minimum validation error. Ties are resolved by smaller validation error and
then by the fixed candidate order. This definition applies to thresholds and
learning rates alike; it does not assume that increasing a DA rate changes
sparsity in a particular direction.

Sites transmit squared-residual sums, fourth-power sums, and sample counts
for each candidate. These give the pooled mean squared error and its ordinary
observation-level standard error. Candidate support sizes are also recorded.
The one-standard-error rule is a model-selection heuristic. It is not a
confidence statement about the selected variables. True coefficients and
test responses are used only for evaluation, never for selection.

The comparison gives the methods the same data, base penalty, validation
rules, and resource limits. ST, P, and DA each search five settings, but
thresholds and learning rates have different effects. Equal candidate counts
therefore do not make the model families equally flexible. Reporting both
selection rules helps show whether a result depends on a particular tuning
preference.

%% file: sections/design.tex
\section{Simulation design and resource limits}
\label{sec:design}

\subsection{Data generation}

The baseline has 800 training observations, 600 predictors, and 30 nonzero
coefficients. We select the nonzero positions at random. Their signs are
equally likely to be positive or negative, and their magnitudes are uniform
on $[0.4,1.2]$. The coefficient vector is shared across sites. Predictor
rows are Gaussian with unit marginal variances and correlation
$\operatorname{Cov}(X_k,X_l)=\rho^{|k-l|}$. Responses follow
$y=X\beta^\star+\varepsilon$, with independent Gaussian errors. The
baseline uses $\rho=0.5$ and signal-to-noise ratio one, where the signal
variance is $(\beta^\star)^\top\Sigma\beta^\star$.

Training observations are split as $(268,268,264)$ across the three
baseline sites. Each replicate also has two independent validation
samples of 200 observations each and an independent test sample of 2000
observations. All three allocations follow the training site proportions.
The first validation sample selects the base penalty; the second selects
each procedure's candidate. The test sample measures prediction error
after both choices are complete. Intercepts are zero and are not fitted.
The generated matrices are used without additional centering or scaling.

We vary one baseline feature at a time in the first eleven scenarios.
The twelfth scenario changes site distributions together, as specified in
Table~\ref{tab:design}. Each scenario contains 50 independent replicates,
giving 600 generated datasets. All methods and epoch counts within a
replicate use exactly the same data. Seeds differ between scenarios.
This design broadens the original study while keeping individual
comparisons interpretable; it is not a full factorial experiment.

\begin{table}[htbp]
\centering\small
\caption{The twelve simulation scenarios. Unlisted settings equal the baseline.
All scenarios have 800 training observations and 30 nonzero coefficients.}
\label{tab:design}
\begin{tabular}{p{0.29\textwidth}p{0.62\textwidth}}
\toprule
Scenario & Setting\\
\midrule
Baseline & $p=600$, $K=3$, $\rho=0.5$, SNR $=1$; sizes $(268,268,264)$\\
Independent predictors & $\rho=0$\\
Strong correlation & $\rho=0.8$\\
Two sites & $K=2$, sizes $(400,400)$\\
Five sites & $K=5$, 160 training observations per site\\
Ten sites & $K=10$, 80 training observations per site\\
Weak signal & SNR $=0.25$\\
Strong signal & SNR $=4$\\
Fewer predictors & $p=300$\\
More predictors & $p=1000$\\
Unequal site sizes & $(640,120,40)$\\
Different site distributions & Correlations $(0,0.4,0.7)$; predictor scales
$(1,1.4,0.7)$; noise multipliers $(1,1.5,0.8)$\\
\bottomrule
\end{tabular}
\end{table}

In the different-distributions scenario, each site's noise standard
deviation is its population signal standard deviation times its stated
noise multiplier. Validation and test observations retain that site's
distribution and scale. Test prediction therefore targets the same
site-proportion mixture as the training population. This scenario varies
predictor and noise distributions; the true coefficients remain common.

At site $j$, the first validation sample selects a penalty from 30
log-spaced values between $\lambda_{\max,j}$ and
$10^{-3}\lambda_{\max,j}$, where
$\lambda_{\max,j}=\|X_j^\top y_j\|_\infty/m_j$.
We use the sample-size-weighted mean of these selected penalties as the
common $\lambda$ for every method. This keeps the comparison connected to
the original locally tuned workflow while holding the objective fixed
within each replicate. Changing the number or size of sites can also
change this inherited penalty. Site-count comparisons therefore describe
the complete procedure, including its penalty rule.

\subsection{Two resource questions}

We use $E\in\{1,5,20\}$ and budget values $B\in\{30,100\}$.
Let $q$ be the number of training paths: $q=1$ for CD and ST and $q=5$
for P and DA. One path-round uploads $8Kp$ bytes. Every procedure receives
the same total upload limit
\[
 U_B=8KpB+144K\quad\text{bytes}.
\]
The last term accommodates setup and validation scalars. CD has one
validation candidate; the other methods have five. Each site transmits
its sample size and selected base penalty, and DA additionally transmits
its local largest eigenvalue. Each candidate requires three validation
scalars per site. All these uploads are counted, and the exact ledger is
included with the results.

The \emph{upload-only comparison} gives each path
$R=\lfloor B/q\rfloor$ rounds. It asks what each procedure achieves
when communication limits training. Extra epochs are allowed to increase
local work.

The \emph{joint comparison} also caps local steps at each site, summed
over all candidate paths, at $5B$, independently of $E$. It gives each path
\[
 R=\min\left\{\left\lfloor B/q\right\rfloor,
              \left\lfloor5B/(qE)\right\rfloor\right\}
\]
rounds. It asks what changes when local computation is also limited.
At $B=100$, CD and ST receive $(100,100,25)$ rounds for
$E=(1,5,20)$; P and DA receive $(20,20,5)$ rounds per candidate.
All runs stop at these prescribed rounds. They do not stop early on a
small update or use validation outcomes to extend training.

For the joint comparison, one CD sweep and one direct full-gradient
evaluation are each charged $4m_jp$ leading dense arithmetic operations
at site $j$. This is an explicit approximation: zero-coordinate shortcuts,
data access, solver overhead, and vectorization affect actual cost. The
common training allowance is therefore $4mp(5B)$ leading operations.
We report actual elapsed times separately. These include common penalty
selection, DA rate setup where applicable, candidate training, and final
candidate evaluation. Offline calculations involving the true support or
pooled reference are excluded from procedure timing. The shared setup and
validation work are outside the training-operation allowance and are
reported separately. These experiments impose common limits; they do not
assert identical hardware execution time or identical budget utilization.

Download traffic is recorded separately, including model broadcasts and
candidate delivery. The main communication limit concerns uploads.
Vectors use dense double-precision storage, including zero entries.
No compression or operational network overhead is simulated.

\subsection{Outcomes and numerical checks}

We record true-support precision, recall, and F1; test mean squared error;
coefficient error; and the gap from a numerically converged pooled Lasso
at the same $\lambda$. A coefficient is active when its magnitude exceeds
$10^{-4}$. Sensitivity analyses use $10^{-6}$ and $10^{-2}$ for evaluating
the same selected fits. They do not retune the methods. An empty selected
model has precision and F1 zero.

We also record the infinity norm of
\[
 b-\soft\!\left(b-\sum_jw_j\nabla f_j(b),\lambda\right).
\]
A zero value is equivalent to the pooled Lasso optimality conditions.
This diagnostic helps distinguish a stationary averaging iteration from
a solution of the specified objective.

Monte Carlo standard errors are replicate standard deviations divided by
$\sqrt{50}$. Differences between methods or epoch counts are calculated
within replicate before averaging. Intervals in the contrast plots are
the paired mean plus or minus 1.96 Monte Carlo standard errors; they
describe simulation precision. We retain every replicate and report the
full result grid in the supplement and accompanying data files.

The CD implementation uses cyclic scikit-learn Lasso with intercept
fitting disabled and a fixed number of sweeps. Independent small-problem
checks compare its output with explicit coordinate updates. The direct
gradient DA implementation is checked against the existing Gram-matrix
implementation of the same update equations. Software versions, code
hashes, seeds, candidate results, and reference-solver diagnostics are
recorded with each computation batch.

%% file: sections/results.tex
\section{Results}
\label{sec:results}

\subsection{The matched comparison favors DA for objective accuracy}

Table~\ref{tab:matched-main} reports the baseline and the scenario with
different site distributions at $B=100$ under both resource limits.
The primary minimum-validation-error rule gives little benefit from
thresholding in the baseline. At $E=5$, mean F1 is 0.261 for CD and
0.268 for both ST and P. Their mean test errors are also similar.
By comparison, DA has F1 0.723 and an objective gap of 0.0067, against
1.644 for ST and P. The paired DA-minus-ST F1 difference is 0.455
(Monte Carlo standard error 0.012).

The same broad pattern appears when site distributions differ. At
$E=5$, DA has F1 0.559, compared with 0.273 for ST and 0.272 for P.
Its mean test error is 53.62, against 56.32 for both modifications.
The paired DA-minus-ST F1 difference is 0.287 (0.023).
Thus the favorable results for DA are not confined to identical site
distributions.

\begin{table}[htbp]
\centering
\caption{Matched comparison at $B=100$ with upload and iterative-work
limits. Candidates minimize independent validation MSE. Entries are
means over 50 replicates, with Monte Carlo standard errors in parentheses.
The objective gap uses the pooled Lasso at the same penalty.}
\label{tab:matched-main}
\input{tables/main_matched}
\end{table}

The full comparison also contains clear exceptions for variable selection.
Across the twelve scenarios and three epoch counts in this resource panel,
DA has a smaller mean relative objective gap than ST and P in all 36
comparisons. Its mean F1 is higher in 30 comparisons and lower in six.
The six unfavorable comparisons occur at all three epoch counts in the
ten-site and weak-signal scenarios. At $E=5$, the paired DA-minus-ST
F1 differences are $-0.126$ (0.030) and $-0.069$ (0.021), respectively.
The corresponding P comparisons are almost identical.

These exceptions help explain what objective accuracy can establish.
The common penalty is inherited from local validation, and the pooled
Lasso itself selects very few variables in these two scenarios. Its
mean active-set sizes are 2.80 and 4.96, although the true support has
30 variables; its F1 values are only 0.145 and 0.212. DA can therefore
approach the specified objective accurately without recovering the true
support well. This is a limitation of the complete penalty-and-fitting
procedure. It does not identify site count alone as the cause, because
site count also changes penalty calibration.

The smaller budget reveals exceptions for objective accuracy as well.
At $B=30$, DA has a larger mean relative objective gap than the other
three methods in six cells: the unequal-size and strong-signal scenarios,
each at $E=1$ in both panels and at $E=20$ under joint limits.
For example, with strong signal at $E=1$, its mean relative gap is
0.261, compared with 0.091 for ST. DA's advantage at the larger
budget is therefore not a guarantee under very short training runs.

\subsection{More epochs help some outcomes and hurt others}

\begin{figure}[htbp]
\centering
\includegraphics[width=\textwidth]{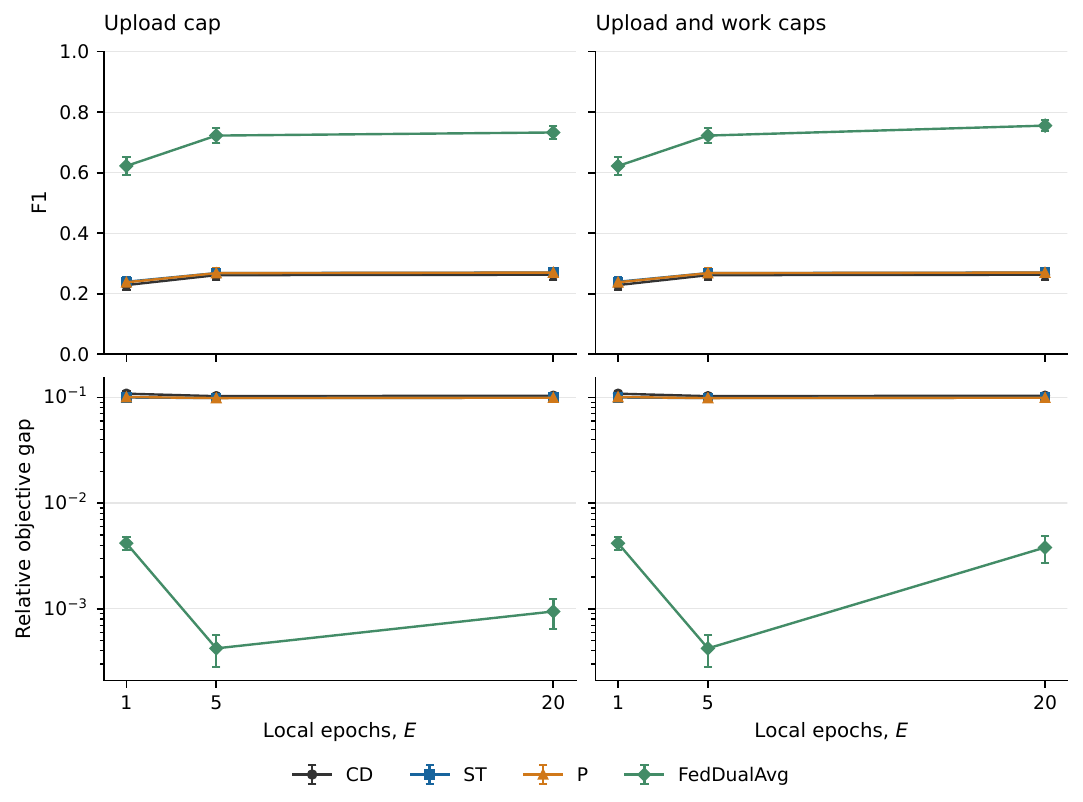}
\caption{Baseline results at $B=100$ under minimum-validation-MSE
selection. The panels distinguish communication limits from joint
communication and iterative-work limits. Relative objective gaps divide
the objective gap by $\max\{1,|F(\hat\beta_F)|\}$.
Error bars show mean $\pm1.96$ Monte Carlo standard errors.}
\label{fig:baseline}
\end{figure}

Figure~\ref{fig:baseline} separates upload-only limits from joint limits.
For CD, ST, and P, most of the baseline F1 improvement occurs between
$E=1$ and $E=5$. Under the joint limits, moving from $E=5$ to $E=20$
raises CD F1 by only 0.0013 (paired standard error 0.0006).
The mean CD objective gap instead rises from 1.707 to 1.724.
For DA, F1 rises from 0.723 to 0.756, while its objective gap rises
from 0.0067 to 0.0603 and test error rises from 26.98 to 27.55.
The preferred epoch count therefore depends on the outcome.

Budget utilization matters when interpreting these changes. At $B=100$,
each workflow uses 100 step equivalents at $E=1$ and 500 at $E=5$ or
$E=20$ under the joint limits. The first comparison therefore allows
more executed training work. Comparing $E=5$ with $E=20$ holds this
leading arithmetic count fixed while allocating it across fewer rounds.
The larger epoch count then uses about one quarter of the training
uploads. Under upload-only limits, $E=20$ instead receives four times
the training work of $E=5$. These are different practical questions.

\begin{figure}[htbp]
\centering
\includegraphics[width=\textwidth]{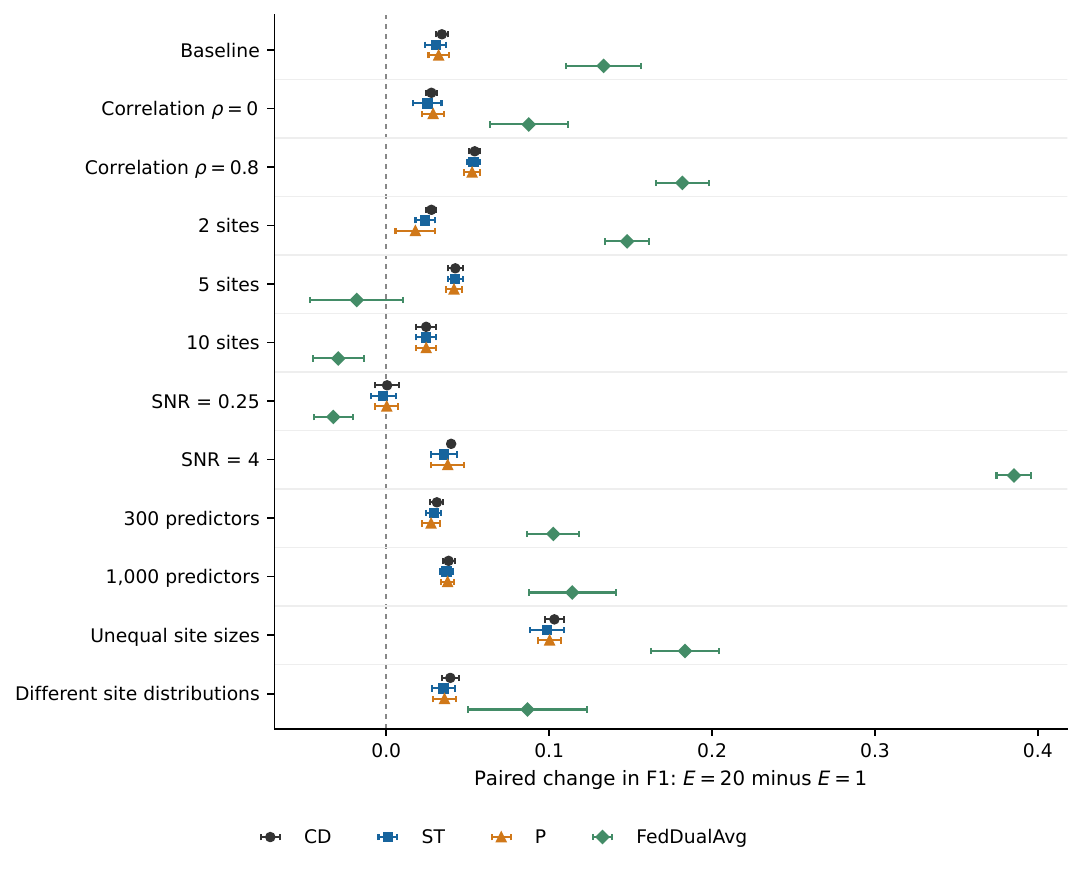}
\caption{Paired changes in true-support F1 from $E=1$ to $E=20$ at
$B=100$ under joint limits and minimum-validation-MSE selection.
Each point averages 50 within-dataset differences. Intervals are
mean $\pm1.96$ paired Monte Carlo standard errors and are not adjusted
for multiple comparisons. Equal resource ceilings do not imply equal
executed work: $E=20$ uses more training steps than $E=1$ here.}
\label{fig:epoch-contrasts}
\end{figure}

Figure~\ref{fig:epoch-contrasts} shows that epoch sensitivity also changes
across scenarios. For CD, the paired F1 increase from $E=1$ to $E=20$
is 0.034 (0.002) in the baseline, 0.054 (0.002) with strong correlation,
and 0.103 (0.003) with unequal site sizes. With weak signal the change
is only 0.001 (0.004). At fixed leading work, the $E=20$ versus $E=5$
CD change is 0.030 (0.002) with unequal sizes, compared with 0.0013
(0.0006) in the baseline. Additional local sweeps are therefore more
consequential in some of the tested designs.

For DA, the $E=20$ versus $E=1$ F1 change is 0.385 (0.005) with
strong signal, but $-0.029$ (0.008) with ten sites and $-0.0325$
(0.0061) with weak signal. The latter decreases also occur when
comparing $E=20$ with $E=5$: $-0.018$ (0.006) and $-0.020$
(0.005). These findings show why a single epoch schedule cannot be
justified by the baseline alone. Three tested values reveal sensitivity;
they do not locate a general optimum.

\subsection{Thresholding results depend on the selection rule}

The secondary one-standard-error rule changes the assessment of ST
and P substantially. In the baseline at $B=100,E=5$ under joint limits,
ST F1 rises from 0.268 to 0.496. The paired increase is 0.228 (0.015),
but test MSE rises from 29.07 to 30.91. P shows a similar trade-off.
Under minimum-error selection, ST and P choose the zero threshold in
43 of these 50 baseline replicates. Their close primary results thus
reflect a frequent validation preference for leaving the fit unchanged.

The one-standard-error rule is not consistently beneficial. For ST,
its paired F1 changes are $-0.170$ (0.028) with ten sites, $-0.264$
(0.014) with weak signal, and $-0.081$ (0.033) with different site
distributions. ST and P select entirely zero models in 76\%, 98\%,
and 46\% of these replicates, respectively. This rule can remove
false positives, but it can also remove nearly all true variables.
Supplement S1 reports the sensitivity for every scenario rather than
only the favorable baseline case.

Under the primary rule, changing the numerical activity cutoff from $10^{-4}$ to $10^{-6}$
changes mean F1 by less than 0.001 in every method--scenario cell
at the same $B=100,E=5$ comparison. Raising it to $10^{-2}$ has
larger effects: CD mean F1 increases by as much as 0.074.
Small averaged coefficients therefore matter to support summaries.
These checks evaluate the existing fits; they do not select a new
threshold using the true support.

\subsection{Resource and numerical diagnostics}

The supplement reports uploads, downloads, training steps, and elapsed
times; the result files also separate penalty setup, rate setup,
training, and validation time. The shared penalty search is substantial:
its median leading-operation count is 6.41 times the iterative count
at $B=100,E=5$. The joint cap should therefore be read as an allowance
for iterative training, not for the complete computation. DA candidate
columns are vectorized, whereas P paths are processed separately;
elapsed-time comparisons describe this implementation.

All 600 pooled reference fits converged, and all 2,200 selected local
penalty fits met the stated tolerance. Some unselected small-penalty
fits reached the initial iteration cap. Extending all affected-scenario
paths from 2,000 to 20,000 iterations resolved those warnings and
changed no selected penalties or reported results. The source package
retains both the original diagnostics and this sensitivity check.

%% file: tables/main_matched.tex
\begingroup\small
\setlength{\tabcolsep}{3.4pt}
\begin{tabular}{lrlll}
\toprule
Method & $E$ & F1 & Test MSE & Objective gap\\
\midrule
\multicolumn{5}{l}{\textit{Baseline: upload and work caps}}\\
CD & 1 & 0.229 (0.008) & 28.41 (0.48) & 1.803 (0.042)\\
CD & 5 & 0.261 (0.008) & 29.03 (0.49) & 1.707 (0.044)\\
CD & 20 & 0.263 (0.008) & 29.08 (0.50) & 1.724 (0.045)\\
\addlinespace[2pt]
ST & 1 & 0.239 (0.008) & 28.45 (0.48) & 1.667 (0.050)\\
ST & 5 & 0.268 (0.008) & 29.07 (0.49) & 1.644 (0.046)\\
ST & 20 & 0.270 (0.008) & 29.12 (0.50) & 1.656 (0.045)\\
\addlinespace[2pt]
P & 1 & 0.238 (0.008) & 28.44 (0.48) & 1.679 (0.043)\\
P & 5 & 0.268 (0.008) & 29.07 (0.49) & 1.644 (0.046)\\
P & 20 & 0.270 (0.008) & 29.12 (0.50) & 1.656 (0.045)\\
\addlinespace[2pt]
FedDualAvg & 1 & 0.622 (0.015) & 27.75 (0.48) & 0.0676 (0.004)\\
FedDualAvg & 5 & 0.723 (0.013) & 26.98 (0.49) & 0.0067 (0.00099)\\
FedDualAvg & 20 & 0.756 (0.009) & 27.55 (0.50) & 0.0603 (0.007)\\
\midrule
\multicolumn{5}{l}{\textit{Different site distributions: upload and work caps}}\\
CD & 1 & 0.226 (0.006) & 56.05 (1.05) & 3.161 (0.095)\\
CD & 5 & 0.262 (0.006) & 56.21 (1.04) & 2.496 (0.071)\\
CD & 20 & 0.266 (0.006) & 56.31 (1.04) & 2.513 (0.073)\\
\addlinespace[2pt]
ST & 1 & 0.240 (0.007) & 56.17 (1.05) & 2.928 (0.110)\\
ST & 5 & 0.273 (0.007) & 56.32 (1.04) & 2.345 (0.084)\\
ST & 20 & 0.275 (0.007) & 56.42 (1.04) & 2.371 (0.085)\\
\addlinespace[2pt]
P & 1 & 0.240 (0.007) & 56.18 (1.05) & 2.952 (0.109)\\
P & 5 & 0.272 (0.007) & 56.32 (1.04) & 2.344 (0.084)\\
P & 20 & 0.275 (0.007) & 56.42 (1.04) & 2.371 (0.085)\\
\addlinespace[2pt]
FedDualAvg & 1 & 0.489 (0.018) & 54.35 (1.06) & 0.0969 (0.011)\\
FedDualAvg & 5 & 0.559 (0.021) & 53.62 (1.09) & 0.0591 (0.008)\\
FedDualAvg & 20 & 0.576 (0.025) & 55.11 (1.12) & 0.380 (0.044)\\
\midrule
\bottomrule
\end{tabular}
\endgroup

%% file: sections/discussion.tex
\section{Discussion}
\label{sec:discussion}

Local epochs control how much fitting occurs before information is
combined. They do not by themselves ensure that coefficient averaging
solves the pooled Lasso. The orthogonal calculation makes this distinction
exact: one sweep already completes each local fit, while averaging can
retain variables selected at only one site. The correlated construction
adds a different point. Increasing the epoch count speeds convergence
per round to the averaging fixed point, yet the limiting objective gap can first
decreases and then increases. Its explicit minimizing epoch count
describes this construction, rather than a general tuning prescription.

The matched experiments clarify the practical value of the thresholding
modifications. ST is inexpensive when a CD trajectory is already
available, because its candidate models share that trajectory. P pays
for separate trajectories and offers little additional benefit in the
main baseline comparison. Under the primary minimum-error rule, both
modifications leave a large objective gap relative to DA. Their stronger
baseline support results under the one-standard-error rule come with
worse prediction, and the same rule performs poorly in several other
scenarios. Thresholding should therefore be evaluated as a specific
model-selection choice, with its cost and adverse cases reported.

For accurately minimizing the common penalized objective, DA is the
stronger procedure at the larger budget; some very short runs favor the
other methods. The larger-budget conclusion does not extend
automatically to recovering the true variables. In the ten-site and
weak-signal scenarios, the inherited penalty produces a pooled target
that omits most true variables. A more accurate solver cannot correct
that statistical target simply by reducing its optimization error.
Practical evaluation should state whether the priority is objective
accuracy, prediction, or variable selection, and assess the penalty rule
alongside the optimization procedure.

Several limits remain. The new study uses Gaussian linear models,
a shared coefficient vector, fixed total sample size and sparsity,
and 50 replicates per scenario. The distributional sensitivity changes
predictor and noise distributions, not the regression coefficients.
Most scenarios change one feature at a time, so interactions are not
mapped. Changing site count also changes the common penalty and the
absolute byte allowance. The five-setting grids equalize candidate
counts, but not model flexibility or optimal tuning effort. The
one-standard-error rule uses a pooled observation-level error estimate;
with different site distributions it is not a stratified standard error.
The arithmetic cap approximates iterative work, and the timing study
does not simulate a deployed network. Finally, the exact two-variable
construction keeps support unchanged and cannot explain every
high-dimensional support pattern.

These limits suggest focused extensions: jointly tune the common penalty
and solver parameters, examine more epoch counts at fixed executed work,
and study coefficient differences across sites. The present evidence
already supports a narrower, useful conclusion. Epoch choice, sparse
averaging, and penalty selection have distinct effects. Measuring them
separately makes both improvements and failures easier to interpret.